\documentclass[pre,reprint,longbibliography,showpacs,twocolumn]{revtex4-2}

\usepackage{amssymb}
\usepackage{graphicx}
\usepackage{adjustbox}
\usepackage{dcolumn}
\usepackage{array}
\usepackage{amsmath}
\usepackage{amssymb}
\usepackage{units}
\usepackage{xcolor}
\usepackage{bm}
\usepackage{setspace}
\usepackage{color}
\usepackage{textcomp}
\usepackage{dcolumn}
\usepackage{booktabs}
\usepackage{upquote}
\usepackage[driverfallback=dvipdfm]{hyperref}
\usepackage[english]{babel}
\usepackage{xurl}

\newcolumntype{d}[1]{D{.}{.}{#1}}
\newcommand\mc[1]{\multicolumn{1}{c}{#1}} 
\usepackage[driverfallback=dvipdfm]{hyperref}
\hypersetup{
    colorlinks=true,
    linkcolor=blue,
    filecolor=magenta,
    urlcolor=black,
    citecolor=red,
    }

\begin{document}
\setlength{\abovedisplayskip}{3pt}
\setlength{\belowdisplayskip}{3pt}
\title{Power Laws for the Thermal Slip Length of a Liquid/Solid Interface\\
From the Structure and Frequency Response of the Contact Zone}
\author{Hiroki Kaifu}
\author{Sandra M. Troian}
\email[]{Corresponding author:\\stroian@caltech.edu; www.troian.caltech.edu}
\affiliation{California Institute of Technology\\
T. J. Watson Sr. Laboratories of Applied Physics\\ Pasadena, CA 91125}
\date{November 26, 2025}

\begin{abstract}
The newest and most powerful electronic chips for applications like artificial intelligence generate so much heat that liquid based cooling has become indispensable to prevent breakdown from thermal runaway effects. While cooling schemes like microfluidic networks or liquid immersion are proving effective for now, further progress requires tackling an age old problem, namely the intrinsic thermal impedance of the liquid/solid (L/S) interface, quantified either by the thermal boundary resistance or thermal slip length. While there exist well known models for estimating bounds on the thermal impedance of a  superfluid/metal interface, no analytic models nor experimental data are available for normal liquid/solid interfaces. Researchers therefore rely on non-equilibrium molecular dynamics simulations to gain insight into phonon transfer at the L/S interface. Here we explore correlated order and motion within the L/S contact zone in an effort to extract general scaling relations for the thermal slip length in Lennard-Jones (LJ) systems. We focus on the in-plane structure factor and dominant vibrational frequency of the first solid and liquid layer for 180 systems. When scaled by the temperature of the liquid contact layer and characteristic LJ interaction distance, the data collapse onto two power law equations, one quantifying the reduction in thermal impedance from enhanced in-plane translational order and the other from enhanced frequency matching in the contact zone. More generally, these power law relations highlight the critical role of surface acoustic phonons, an area of focus which may prove more useful to development of analytic models and instrumentation for validating the relations proposed.
\end{abstract}

\maketitle

\section{Introduction}
High performance CPUs and GPUs for power intensive applications such as artificial intelligence and cryptocurrency exchange generate such tremendous heat within such small volumes that chip designers have had to pivot from air to liquid cooling to prevent failure from thermal runaway and consequent deleterious behavior \cite{VM20,R23,RS23}. Liquid cooling has also demonstrated faster clock speeds, higher efficiency, improved performance and better stability in systems ranging from conventional CMOS and superconducting processors to solid-state quantum devices \cite{SD24}. While aqueous liquids are still common, liquid metals and alloys are of growing interest because of their superior thermophysical and other properties such as high thermal and electrical conductivity, high boiling point, high surface tension and low viscosity \cite{L22}. Metallic based liquids can also be transported throughout electronic devices using compact magnetofluid dynamic pumps, which are vibration-free and therefore operate quietly and efficiently.

\begin{figure}[!htb]
\centering
\includegraphics{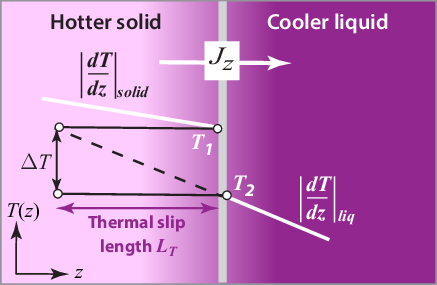}
\vspace{-0.1 in}
\small
\caption{Illustration of the thermal slip length $L_T= \Delta T/|dT/dz|_{liq}$.}
\label{fig:Ldefn}
\end{figure}

Cooling schemes using two-phase cooling in microfluidic networks or direct liquid immersion are currently in use and being refined. However, further progress requires tackling the age-old issue of the intrinsic thermal impedance of any liquid/solid (L/S) interface due to the discontinuity in material properties at the boundary. This impedance is typically quantified by the magnitude of the thermal boundary resistance or the thermal slip length. The latter quantity if the preferred measure in this study to draw analogy with the velocity slip length in hydrodynamic systems. Illustrated in Fig. \ref{fig:Ldefn} is the thermal slip length of a L/S interface for a constant thermal flux $J_z$ propagating in the direction normal to the interface, here oriented along the $\hat{z}$ axis. It is defined by the relation
\begin{equation}
L_T= \frac{\Delta T}{~~~\big|dT/dz \big|_{liq}}~,
\label{eqn:sliplength}
\end{equation}
where $|dT/dz|_{liq}$ is the magnitude of the thermal gradient in the liquid interior, which in this study reduced to a constant that depended on the input variables due to the linearity of the temperature profile. While in macroscopic L/S systems the interface thermal impedance is far smaller than that of the bulk liquid and solid layers and therefore negligible, that is not the case in micro- or nanoscale systems which manifest very large surface to volume ratios.

Equation (\ref{eqn:sliplength}) is just a definition, of course, since all the dependencies and complexities of a given system are buried in the variables $\Delta T$ and $J_z$. Despite decades of effort, the only L/S systems for which there exist predictive models for estimating the thermal boundary impedance are those pertaining to  superfluid helium/metal interfaces. On the experimental side, superfluid/metal systems have special properties which allow accurate measurement of variables such as temperature, pressure, elastic properties of the solid and 1iquid and even the excitation spectrum of phonons in the bulk and at the interface. Even so, the two best known analytic relations for such systems only provide bounds on the magnitude of the thermal boundary resistance (i.e. Kapitza resistance) depending on whether phonon behavior at the interface is predominantly specular or diffusive  \cite{K52,P69,SP89}. There are no such equations for normal L/S systems (i.e. not involving superfluids).

This situation poses a serious problem on a fundamental level when trying to solve for the temperature distribution throughout an L/S system based on the differential equations governing thermal transfer. At macroscopic scales, since the interface boundary resistance is relatively negligible, $\Delta T = T_1 - T_2 \approx 0$ such that the temperature of the solid surface is identical to that of the first liquid monolayer in contact with the substrate - which in this study is called the \emph{contact layer}. In micro- or nanoscale systems however, the boundary values $T_1$ and $T_2$ are unknown. Furthermore, this dilemma cannot be resolved by appealing to experimental data since, aside from the special class of systems mentioned above, there are no adequate experimental probes for measuring $\Delta T$ in normal L/S systems.

\subsection{Analogy with velocity slip length at a L/S interface}

There exists an analogous dilemma in hydrodynamic systems  involving velocity slip at the interface of a liquid and solid in relative motion. Until recently, the velocity boundary condition (BC) needed to solve Cauchy's equation of motion was based on a phenomenological relation known as the Navier slip law \cite{N1823}, wherein the velocity slip length is treated as an unknown constant. Unlike other boundary conditions needed to solve the governing equations for mass, momentum and energy transport, the thermal slip and velocity slip boundary conditions are unique in that they cannot be deduced from conservation laws or considerations of symmetry. For this reason, researchers in both fields have come to rely heavily on non-equilibrium molecular dynamics (NEMD) simulations to uncover correlations among the many system variables and the corresponding slip lengths.

During the past several decades, advances in NEMD simulations have helped reveal many aspects of velocity slip in systems ranging from simple liquids or polymeric fluids to more complex fluids flowing across the surface of smooth or rough, wetting or non-wetting, chemically patterned or textured substrates. An early study revealed that when normalized by key asymptotic variables, the velocity slip length exhibits a distinctive power law dependence on the liquid shear rate \cite{TT97}. That boundary condition has been adopted extensively and further generalized to describe many different systems. Lesser known or emphasized, but equally important, is the fact that the velocity slip length exhibits a strong inverse dependence on the peak value of the in-plane structure factor of the contact layer. This dependence, which hinges on the degree of translational order within the contact layer induced by the substrate potential, has since been verified in many simulations and even validated by analytic models for certain systems \cite{TR90,BB99,PT04,PT05,PT06}.

\subsection{Relevant prior studies and open questions}
The phenomenon of thermal slip at a normal L/S interface has also been investigated extensively by NEMD simulations, which have revealed the influence of system properties such as the wettability of the L/S interface \cite{M97,MK98,OS00,BC03,XE03,WK11}, pressure of the bulk liquid against the solid surface \cite{MP08a,HM17}, temperature of the solid surface \cite{BP08,MK13}, solid surface roughness \cite{WK11}, symmetry of the solid lattice \cite{OT05,TI10,KB24}, thickness of the liquid layer between two solid lattices at different temperatures  \cite{KC08}, width of the L/S density depletion zone \cite{RP16,RP18,LW20} and so on. However, no general relations for predicting the overall magnitude of the  thermal slip length have been developed, in part because of  the difficulty in untangling effects arising from poorly understand interactions among the various system parameters. In this study, we therefore adopt a different approach by focusing on two key measures of \textit{correlated behavior} within the L/S contact zone, defined as the interfacial region spanning the first solid and first liquid layer i.e. contact layer. These two measures, corresponding to the structural order and the dominant vibrational frequency of the contact layer, reveal the influence of the solid substrate potential on the transmission of acoustic phonons across the L/S boundary.

\subsection{Choice of intermolecular potential in NEMD simulations}
\label{sec:potchoice}
The overwhelming majority of computational studies on thermal transport across a L/S interface have utilized the well-known 12-6 Lennard-Jones (LJ) pair potential, which for decades has served as the canonical reference when investigating fundamental mechanisms involving statistical mechanical processes. The LJ potential offers a simple yet accurate description of the balance between attractive and repulsive interactions between neutral particles (i.e. molecules or molecular units with closed-electron shells). This potential, now regarded as the archetype model for efficient and realistic simulations, is also often used as the building block for more complex substances involving bonded interactions. For simple metallic systems such as  FCC metal interfaces, the LJ potential is capable of generating highly accurate material constants with far less  computational effort than embedded atom potentials or density functional calculations \cite{HN08,KH21}.

A key feature of the LJ potential is its general form given by $U = \varepsilon ~ \mathcal{U} (r/\sigma)$, where $\varepsilon$ and $\sigma$ specify characteristic energy and distance scales and $\mathcal{U}$ denotes a universal function of the scaled spatial coordinate $r/\sigma$. According to the law of corresponding states \cite{HR60}, transport coefficients including the thermal diffusivity, molecular diffusivity and kinematic viscosity can be directly mapped from one system to another by simply re-scaling the constants $\varepsilon$ and $\sigma$. For this reason, while many studies including ours are based on the scales and properties of argon, the results are more generally applicable.

\subsection{Motivation for current study}
In a recent study \cite{KB24}, we demonstrated a strong correspondence between the magnitude of the thermal slip length and the motion of liquid particles in the contact layer. In particular, those simulations revealed how the depth and width of the corrugation of the L/S periodic surface potential control the degree of particle localization by repressing in-plane migration and diffusion. We coined that behavior \textit{2D caging} in reference to the well-known 3D caging phenomena leading to glassy behavior in amorphous systems. There we showed that 2D caging enhances thermal transfer out-of-plane between liquid layers, thereby reducing the thermal slip length. Informed by those findings, we wanted to explore the  process in more detail by examining the relation between the thermal slip length and two important measures of the L/S contact zone as quantified by the structural and vibrational characteristics of the contact layer. The results we report here, based on analysis of 180 different L/S systems interacting via LJ potentials, demonstrate the existence of distinct power law relations for the thermal slip length described in the spatial and frequency domain.

\begin{figure}[!htb]
\centering
\includegraphics{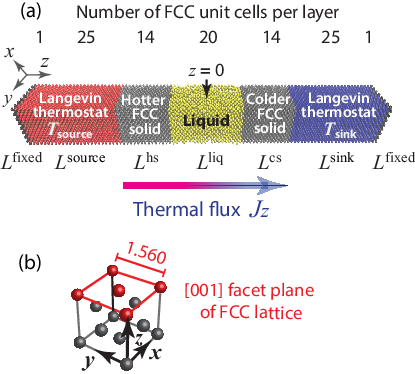}
\vspace{-0.1 in}
\small
\caption{(a) Layered geometry of entire computational cell. Scalings of variables and layer dimensions for the geometry can be found in Tables \ref{tbl:dimlessunits} and \ref{tbl:slablengths}. Coordinate origin $z=0$ was situated at the midplane of the liquid layer. (b) FCC crystal unit cell with lattice constant 1.560 (reduced units) showing [001] facet plane (red). For all runs, the surface normal to the [001] plane was oriented parallel to the thermal flux vector $J_z$.}
\label{fig:schematics}
\end{figure}

\section{Computational Details}
This section outlines details of the NEMD simulations using the open source package LAMMPS, a flexible tool for particle--based simulations of gases, liquid and solids in systems spanning from atomic to macroscopie length scales \cite{P95,TP22}. Additional details pertaining to these simulations can also be found in Ref. [\onlinecite{KB24}].

\begin{table}[!htb]
\small
\setlength{\tabcolsep}{1pt}
\renewcommand{\arraystretch}{1.2}
\begin{tabular}{l l l}
\hline
\hline \\ [-8 pt]
\,Physical quantity & Numerical value
\vspace{0.02 in}\\
\hline \\ [-8 pt]
\,mass & $m^\ast = 6.690 \times 10^{-26}$ kg\\
\,length & $\sigma^\ast = 0.3405 \times 10^{-9}$ m \\
\,energy  & $\epsilon^\ast = 165.3 \times 10^{-23}$ J \\
\,temperature  & $T^\ast= \epsilon^\ast/k_B = 119.8$ $^\circ$K \\
\,time  & $t^\ast=(m^\ast {\sigma^\ast}^2 / \epsilon^\ast)^{1/2}=2.14$ ps\\
\,mass density & $\rho^\ast = m^\ast/(\sigma^\ast)^3$\\
\,pressure & $p^\ast= \epsilon^\ast/(\sigma^\ast)^3 = 0.4187$ MPa\\
\,effective particle diameter  & $\sigma^\ast_{LL} = \sigma^\ast_{SS}=\sigma^\ast$\\
& $\sigma^\ast_{LS}= 0.8, 1.0, 1.2~\sigma^\ast$\\
\,FCC edge length & $a^\ast \!=\! 1.560 \, \sigma^\ast \!=\! 5.312 \times 10^{-10}\,\textrm{m}$\\
\,interaction energy  & $\epsilon_{LL} = \epsilon^\ast$\\
\, & $\epsilon_{LS} = 0.1 - 1.0 ~ \epsilon^\ast$\\
\, & $\epsilon_{SS} = 10 ~\epsilon^\ast$ \\ [+2 pt]
\hline \\ [-10 pt]
\,Variable & Value in scaled units
\vspace{0.02 in} \\
\hline \\ [-8 pt]
\,solid \& liquid particle mass & 1.0 \\
\,LJ repulsive distance & $\sigma_{LL} = \sigma_{SS} = 1.0$\\
\,& $\sigma_{LS} = 0.8, 1.0 ~\textrm{or}~ 1.2$\\
\,FCC edge length & $a=1.560$ \\
\,integration time step & $\Delta t_{int} = 0.002$\\
\,thermostat setting &
$(T_\textrm{source}, T_\textrm{sink})=$\\
\, & $(1.8,0.8), (1.6,1.0) ~\textrm{or}~ (1.4,1.2)$\\
\,LJ interaction energy & $\varepsilon_{LL} = 1.0$\\
\,  & $\varepsilon_{LS} = 0.1, 0.2, \ldots, 0.9, 1.0$ \\
\,  & $\varepsilon_{SS} = 10$ \\
\,bulk liquid density & $\rho_L \approx$ 0.84 \\
\,FCC unit cell density & $\rho_S = 1.0536$ \\ [+4 pt]
\hline
\hline
\end{tabular}
\caption{Symbols, quantities and numerical values used to rescale variables based on the elemental fluid argon \cite{MW49,V67,HE83}. Asterisk superscripts denote dimensional quantities. Boltzmann's constant $k_B = 1.380649 \times 10^{-23}$ J/K.}
\label{tbl:dimlessunits}
\end{table}
\begin{table}[!ht]
\small
\setlength{\tabcolsep}{1 pt}
\renewcommand{\arraystretch}{1.3} 
\begin{tabular}{ l d{4.1}}
\hline
\hline \\ [-12 pt]
Cell sizes (scaled by $\sigma^\ast$) & \mc{~~FCC [001]}
\vspace{0.02 in} \\
\hline \\ [-12 pt]
$\quad L_x$ & 12.48 \\
$\quad L_y$ & 12.48 \vspace{0.02 in} \\
\hline \\ [-12 pt]
$\quad L^\textrm{fixed}$ (1 unit cell per end) & ~~1.56 \\
$\quad L^\textrm{source}$ & 39.00 \\
$\quad L^\textrm{hs}$ & 21.84 \\
$\quad L^\textrm{liq}$ & 31.20 \\
$\quad L^\textrm{cs}$ & 21.84 \\
$\quad L^\textrm{sink}$ & 39.00 \\
\hline \\ [-12 pt]
Total length along $z$ axis & 156.00 \vspace{0.02 in} \\
\hline
\hline
\end{tabular}
\normalsize
\caption{Dimensions of layered system in  Fig.~\ref{fig:schematics}.}
\label{tbl:slablengths}
\end{table}
\vspace{-0.1 in}
\subsection{Model geometry and interaction potentials}
The simulations conducted were based on the rectangular layered structure shown in Fig. \ref{fig:schematics} (a) and (b) describing a quiescent liquid layer sandwiched between two identical crystalline solids modeled by face-centered cubic (FCC) lattices. Each solid was maintained at a constant temperature by direct thermal contact with another FCC lattice acting as a Langevin thermostat. The left thermostat was set to a temperature $T_\textrm{source}$ and the right one to $T_\textrm{sink})$, which naturally generated a constant thermal flux along the $\hat{z}$ axis. Particles in the outermost layer at each end of the cell were affixed in place to prevent sublimation. All solid lattices were oriented with their [001] facet plane parallel to the L/S interface. Since the mass of all liquid and solid particles was set equal to one (in reduced units), the mass density equaled the number density. Periodic boundary conditions were enforced along the $\hat{x}$ and $\hat{y}$ axes.

All particles were made to interact via a truncated and shifted LJ potential given by

\begin{equation}
U_{\textsf{LJTS}} (r_{ij}) =
\begin{cases}
U (r_{ij}) - U (r_\textrm{c})& \text{if } r_{ij} \leq r_\textrm{c}~,\\
0 & \text{if } r_{ij} > r_\textrm{c}
\end{cases}
\label{eqn:LJ1}
\end{equation}
where
\begin{equation}
U (r_{ij})= 4 \, \varepsilon_{ij} \Big[\Big(\frac{\sigma_{ij}}{r_{ij}}\Big)^{12}-
\Big(\frac{\sigma_{ij}}{r_{ij}}\Big)^{6}\Big]~.
\label{eqn:LJ2}
\end{equation}
This form ensures no discontinuity in the intermolecular force despite the interaction cutoff radius $r_\textrm{c}$. Here, subscripts $ij$ denote pairwise interacting particles $i/j = L/L$, $S/S$ or $L/S$, $r_{ij}= |\bm{r}_i - \bm{r}_j|$ is the pairwise separation distance between particles $i$ and $j$, $\varepsilon_{ij}$ is the pairwise interaction energy and $\sigma_{ij}$ is the pairwise distance where $U(r=\sigma_{ij})=0$, also called the effective particle diameter.

The input parameters for each run were $(T_\textrm{source}, T_\textrm{sink})$, $\sigma_{LS}$ and $\varepsilon_{LS}$, the latter range spanning the so-called non-wetting to wetting regimes. The specific choice of pair values  $(T_\textrm{source}, T_\textrm{sink})$ ensured not only that the interior of the liquid layer in all runs remained far from any critical or triple point \cite{HE83,TV16} and also remained close to the average temperature $(T_\textrm{source} + T_\textrm{sink})/2 = 1.3$ and average density $\rho_{bulk} \approx 0.84$. The computational cell was designed with a relatively large liquid layer. Simultaneous measurements at the hotter and colder side therefore helped reveal the influence of contact layer temperature $T_c$. In total then, the geometry helped generate 180 L/S interfaces for analysis.

The majority of NEMD studies in this field report that the crystalline solids are constructed using a harmonic wall-spring model in which particles are closely tethered to sites of a periodic lattice using a Hookean spring force \cite{CB01,BC08,TP22}. Depending on the temperature range and other input variables, this type of construction can dampen or altogether eliminate anharmonic phonons. To prevent this, the particles in the solid layers in this study were made to interact via a strong-binding LJ potential \cite{SB88,GH14,SV16,HK19} with $\varepsilon_{SS}=10$ and $\sigma_{SS} = 1.0$, while the liquid/liquid interaction constants were set to $\varepsilon_{LL}=1.0$ and $\sigma_{LL}= 1.0$. Since the melting (m) temperature of an LJ solid can be estimated from the relation $T_m \simeq \varepsilon_{SS}/2$ \cite{SN07}, the value $\varepsilon_{SS}=10$ ensured that the crystal remained in the solid state for the temperatures generated in this study. Simulations at isothermal conditions for modeling simple face centered cubic (FCC) metals have shown that similar choices of intermolecular constants yield accurate material property values \cite{HN08,KH21}.

The thickness of the two lattices acting as Langevin thermostats was also chosen to exceed that of the unconstrained solid layers in order to avoid spurious reductions in thermal boundary resistance \cite{HM17}. Studies have shown \cite{LK14,HM17} that when the phonon mean free path in the thermal reservoir region satisfies the relation $\Lambda=c_\ell \times \tau_\textrm{damp} \leq 2 L$, where $c_\ell$ is the speed of longitudinal sound waves, $L$ is the reservoir layer thickness and $\tau_\textrm{damp}$ is the Langevin damping constant, phonons are then dissipated before undergoing reflections from the exterior boundary toward the L/S interface. For an FCC crystal, the value $c_\ell$ was estimated from the relation \cite{SN07} $c_\ell = 9.53 \sqrt {\varepsilon_{SS}}$. Therefore, for the parameters in our study, namely $\tau_\textrm{damp}=1$ and
$L_s = L^\textrm{source}=L^\textrm{sink} = 39$, 																							 the inequality $\Lambda=c_\ell \times \tau_\textrm{damp} = 9.53 \sqrt {10} \simeq 30 \leq 2 L_s = 78$ was well satisfied.

\subsection{Thermal flux regulation}
The S/L/S layers were first thermally equilibrated using a canonical ensemble (constant NVT conditions) at a temperature $T=1.3$ using a Nos$\acute{e}$--Hoover thermostat \cite{H85} for a period $t_{eq}=10^5 \Delta t_{int}=200$. That thermostat was then turned off and a Langevin thermostat \cite{SS78} applied to particles in the two solid layers acting as the thermal source and sink to maintain each at a different fixed temperature $[T_\textrm{s}$ by enforcing the Langevin equation (reduced units)
\begin{equation}
\frac{d^2 \bm{r}_i}{dt^2}=-\sum_{i\neq j}
\nabla U_{\textsf{LJST}} \cdot \bm{r}_{ij} -
\frac{1}{\tau_\textrm{damp}}\frac{d\bm{r}_i}{dt}+
\bm{F}_\textrm{stoch}~.
\label{eqn:langevin}
\end{equation}
The damping constant was chosen to be  $\tau_\textrm{damp}=500 ~\Delta t_{int} = 1.0$ and the  magnitude of the normally distributed random force $F_\textrm{stoch}$ set to the value $[T_\textrm{s}/(\tau_\textrm{damp}~\Delta t_{int})]^{1/2}$. The entire system was then stabilized for an additional period $2 \times 10^{5} \Delta t_{int} = 400$ to ensure a steady uniform thermal flux propagating across the S/L/S system. Measurements of various properties were then extracted from particle trajectories following Newton's equation using second order Verlet integration \cite{V67} with a time step $\Delta t_{int}= 0.002$.

The thermal flux across the system was extracted from the relation
\begin{equation}
J_z =\frac{1}{L_x \times L_y}\bigg \langle \frac{E_\textrm{net}(t)}{t}\bigg \rangle ~,
\label{eqn:Jz}
\end{equation}
where $E_\textrm{net}(t)$ is the net energy input during an interval $t$ required for maintaining the set point values  $(T_\textrm{source}, T_\textrm{sink})$. Angular brackets here and elsewhere denote ensemble averaging described  below. It was confirmed that $<E_\textrm{net}(t)>$ increased linearly in time, verifying a constant thermal flux was established. The mean and standard deviation of the thermal gradient $|dT/dz|$ within the bulk of the liquid and solid regions were extracted from linear least squares fits.

It has been reported that application of high pressure to a liquid can lead to a 3 to 4 fold reduction in the thermal impedance of the L/S interface \cite{HM17} stemming from ultra dense packing of particles against the solid wall. To eliminate this effect from our study, we checked the typical magnitudes of the virial and kinetic contributions to the pressure $p$ in the liquid interior. As an example, for $(T_\textrm{source}, T_\textrm{sink}) = (1.6, 1,0)$ and $\sigma_{LS} = 1.0$, the virial contribution for $\varepsilon_{LS} = 0.1$ was $p=2.72 \pm 0.03$ and for $\varepsilon_{LS} = 1.0$ was $p=2.54 \pm 0.02$. The kinetic contribution to the pressure $p=1.5$ remained constant for all runs since the average temperature in the liquid interior was designed to remain near $T=1.3$. The total pressure within the liquid interior was found always to be in the low single digits, far smaller than the pressures needed to induce a sizeable reduction in thermal impedance from packing effects alone.

\subsection{Measurements extracted}
The geometry in Fig. \ref{fig:schematics} allowed  simultaneous measurement of various quantities from the hotter and colder side of the liquid layer while subject to the same thermal flux. In this study, key measurements extracted from the hotter and colder interface included the temperature $T_c$ and peak density $\rho_c$ of the contact layer, width of the liquid density depletion layer $\delta_{LS}$ (i.e. separation distance between the peak density of the contact layer and first solid layer), thermal gradient within the interior of the liquid and solid layers, temperature drop $\Delta$ across the L/S interface, thermal slip length $L_T$, maximum value of the in-plane static structure factor of the contact layer $S^\parallel_\textrm{max}$, and the frequencies $\nu_S$ and $\nu_L$ corresponding to the peak values in the phonon density of states for the first solid (S) layer and contact (L) layer.

\subsubsection{Ensemble averaging of stationary quantities}
After a constant thermal flux was established, trajectory data were sampled at intervals $500 ~\Delta t_{int} = 1.0$  for a total period $t_\textrm{total}= 5 \times 10^6 \Delta t_{int}= 10^4$. The sampling interval equalled the approximate decay interval of the velocity auto-correlation function of particles in the contact layer. These data strings were divided into ten non-overlapping segments for ensemble averaging.

The density and temperature distributions along the $\hat{z}$-axis were obtained by dividing the S/L/S partitions into non-overlapping bins of volume $L_x \times L_y \times \Delta z_\textrm{bin}$. A slender bin width of $\Delta z_\textrm{bin} = 0.016$ was used to capture fine details of the oscillations in the liquid layer near solid surfaces. The average density in each bin was estimated to be $\rho_\textrm{bin}=\langle N_\textrm{bin} \rangle /V_\textrm{bin}$, where $N_\textrm{bin}$ represents the average number of particles in a bin. The thickness of the contact layer was defined to be the distance separating the adjacent minima of the first and largest oscillation in the liquid density profile near the solid surface; this peak density of the contact layer is denoted $\rho_c$. It was confirmed that the speed of particles in both the contact and first solid layer conformed to a Maxwell-Boltzmann distribution, thereby reflecting a state of \textit{local thermal equilibrium}. The average temperature in each bin (based on a bin width $\Delta z_\textrm{bin} = 0.785$) was therefore computed from the equipartition relation
\begin{equation}
T_\textrm{bin} = \bigg \langle \frac{1}
{3\,N_\textrm{bin}}
\sum_{i}^{N_\textrm{bin}} \bm{v}^2_i \bigg \rangle ~,
\label{eqn:CLtemp}
\end{equation}
where $\bm{v}_i$ denotes the 3D velocity vector of particle $i$.

The temperature drop across the L/S interface was obtained by extrapolation of the linear temperature profile (confirming thermal conduction) within the solid and liquid layer toward the L/S interface. The value $\Delta T$ represents the temperature drop at the midpoint of the distance separating the peaks in density of the first solid and contact layer. This separation distance is also sometimes called the depletion layer thickness $\delta_{LS}$. The thermal slip length was then obtained from the relation
\begin{equation}
L_T= \bigg \langle \frac{\Delta T}{~~~\big|dT/dz \big|_{liq}}\bigg \rangle~.
\label{eqn:sliplength}
\end{equation}

The degree of long range translational order within the contact (c) layer was quantified by the in-plane static structure factor \cite{FN1new}
\begin{equation}
S^\parallel_c(\bm{k})=\bigg<\frac{1}{N^2_c}\sum_{p=1}^{N_c}\sum_{q=1}^{N_c}
\exp \big[i \bm{k} \cdot (\bm{r}_p - \bm{r}_q ) \big] \bigg>~,
\label{eqn:S(k)}
\end{equation}
where $\parallel$ signifies the planar coordinates $\bm{r}=(x,y)$ and wave numbers $\bm{k}=(k_x,k_y)$ for the total number of particles in the layer $N_c$. Equation (\ref{eqn:S(k)}) was normalized to span the range $0 \leq S^\parallel_c(\bm{k}) \leq 1$. We confirmed that the global maximum of Eq. (\ref{eqn:S(k)}), denoted by $S^\parallel_\textrm{max}$, always coincided with the set of smallest reciprocal lattice vectors of the [001] facet plane of the FCC solid lattices.

\subsubsection{Ensemble averaging of time-dependent  quantities}
Measurements of the velocity autocorrelation function were collected over a total period $t_\textrm{total}= 1.5 \times 10^6 \Delta t_{int}= 3 \times 10^3$, then divided into three non-overlapping equal time blocks with initial times  $t^B_o = 0$, $10^3$ and $2 \times 10^3$. Velocities in each block were sampled at intervals $10 \times \Delta t_{int}=0.02$, which generated a sequence of autocorrelation values spanning the interval $t_f - t_o$ for $t_o=t^B_o + (0, 10, 20, \ldots, 475,000) \times 0.02$. Since particles in the first solid layer remained in that layer throughout, the final sampling time was set to $t_f=50$. Since particles in the contact layer could exit and re-enter that layer, a different strategy was needed to establish an appropriate interval for evaluating correlations. Autocorrelation data were therefore restricted to those subset of particles in the contact layer $N_L(t_o,t_f) \geq 10$, confirmed to remain in that layer throughout the interval $t_f - t_o$. In all cases, we ensured that this interval exceeded the velocity autocorrelation decay time by at least an order of magnitude.

The phonon density of states per particle $\mathcal{D}(\nu)$, representing the spectrum of normal mode vibrations, was computed from the relation
\cite{BW83,LG03}
\begin{equation}
\mathcal{D}(\nu)\!=\!
\Bigg\langle \!\frac{4}{N_LT_L} \!\int^{t_f}_{0} \! \sum_{j=1}^{N_L}\bm{v}_j(t_o\!+\!t) \cdot \bm{v}_j(t_o) \cos(2\pi \nu t)\,dt \!\Bigg\rangle^{\!\!\!B}_{\!\!\!t_o},
\label{eqn:DOS}
\end{equation}
where $T_L$ denotes the temperature of the first solid layer or contact layer, as appropriate. Equation (\ref{eqn:DOS}) was normalized to satisfy the equipartition relation $\int_0^\infty \mathcal{D}(\nu)\, d\nu = 3$, reflecting three degrees of freedom for vibrational motion. Since different initial times $t_o$ led to different final times $t_f$, the smallest value $t_f$ within each time block was used to estimate the mean value for that block. The smallest value $t_f$ of all three blocks then used to compute the final block (B) average for $\mathcal{D}(\nu)$. The notation $< \cdot >^B_{t_o}$ signifies the ensemble average over initial times $t_o$ followed by the three-block average.

\section{Simulation Results}
\subsection{Behavior of thermal flux $J_z$}
\begin{figure}[!htb]
\centering
\includegraphics{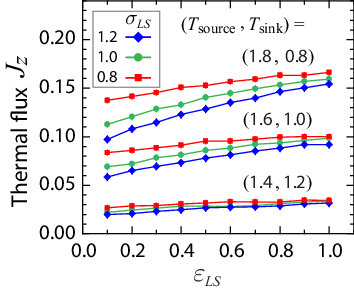}
\vspace{-0.1 in}
\small
\caption{Increase in thermal flux $J_z$ with increasing $T_\textrm{source}-T_\textrm{sink}$, increasing $\varepsilon_{LS}$ and decreasing $\sigma_{LS}$. Connector lines are a guide to the eye.}
\label{fig:flux}
\end{figure}

Shown in Fig. \ref{fig:flux} is the increase in thermal flux $J_z$ across the S/L/S system for larger values  $T_\textrm{source} - T_\textrm{sink}$ and  $\varepsilon_{LS}$ and smaller values $\sigma_{LS}$. For constant $\varepsilon_{LS}$, the highest thermal flux is achieved by the largest differential $T_\textrm{source}-T_\textrm{sink}$ and smallest value $\sigma_{LS}$. For the smallest applied differential $(1.4, 1.2)$, $J_z$ is relatively insensitive to $\sigma_{LS}$ and far less sensitive to $\varepsilon_{LS}$. Overall, the larger the differential $T_\textrm{source} - T_\textrm{sink}$, the stronger the influence of $\varepsilon_{LS}$ and $\sigma_{LS}$ on the thermal flux.

\subsection{Behavior of temperature jump $\Delta T$}
\begin{figure}[!htb]
\centering
\includegraphics{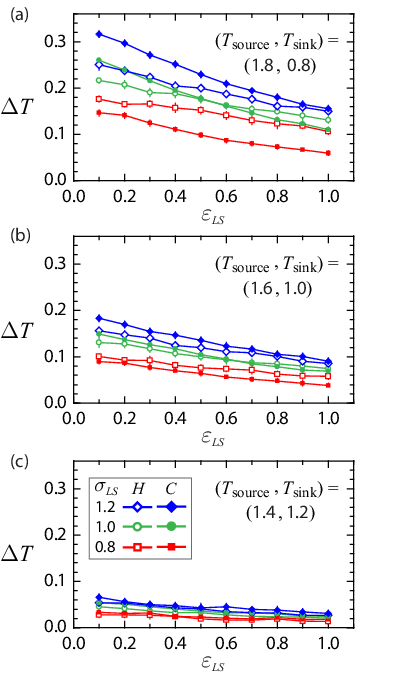}
\vspace{-0.1 in}
\small
\caption{(a) - (c) Reduction in the temperature jump $\Delta T$ at the hotter (H) and colder (C) L/S interface for smaller differential $T_\textrm{source} - T_\textrm{sink}$, smaller $\sigma_{LS}$ or larger  $\varepsilon_{LS}$. Connector lines are a guide to the eye.}
\label{fig:Tjump}
\end{figure}

Shown in Fig. \ref{fig:Tjump} is the reduction in the temperature jump $\Delta T$ at the hotter (H) and colder (C) L/S interface for smaller differential $T_\textrm{source} - T_\textrm{sink}$, smaller $\sigma_{LS}$ or larger $\varepsilon_{LS}$.
Further inspection of the data Fig. \ref{fig:Tjump} reveals some unexpected features. (The tabulated entries in Ref. \onlinecite{KT25} evidence the trends reported below in more detail.) For example, when $\sigma_{LS}=1.2$, the temperature drop at the colder L/S interface is always larger than that at the hotter interface irrespective of the value $\varepsilon_{LS}$. However, the opposite behavior is observed for $\sigma_{LS}=0.8$, as clear from  Figs. \ref{fig:Tjump} (a) and (b). For the systems in Fig. \ref{fig:Tjump} (c), the tabulated entries \cite{KT25} confirm that the behavior resorts back to larger thermal jumps at the colder interface. For the intermediate case $\sigma_{LS}=1.0$, the thermal jump at the colder side can be larger or smaller that at the hotter side depending on the value $\varepsilon_{LS}$; this transition occurs near  $\varepsilon_{LS} = 0.6$. Another interesting observation is that the smallest overall temperature drop occurs at the hotter interface for values $(T_\textrm{source}, T_\textrm{sink})=(1.4, 1.2)$, $\sigma_{LS}=0.8$ and $\varepsilon_{LS}=1.0$. Based on considerations of kinetic energy and collision frequency between L and S particles, one might have expected the smallest temperature drops in the hottest layers generated with the setting $T_\textrm{source}=1.8$, but that is not the case. This then illustrates that $\Delta T$ is influenced by a number of variables including s $T_\textrm{source}, T_\textrm{sink}$, $\varepsilon_{LS}$ and $\sigma_{LS}$ and so its behavior cannot necessarily be intuited \textit{a priori}.

This introduces another misconception which occasionally  creeps into the literature, namely that the higher the contact density $\rho_c$, the smaller must be the temperature drop $\Delta T$ because of the more numerous L/S collisions per unit area \cite{MP08b,PK13}. While the data \cite{KT25} confirm that for certain values of $(T_\textrm{source}, T_\textrm{sink})$ and $\sigma_{LS}$, an increase in $\varepsilon_{LS}$ does cause an increase in $\rho_c$ and decrease in $\Delta T$. However, the data in Fig. \ref{fig:Tjump} show when $\varepsilon_{LS}$ is held constant and $\sigma_{LS}$ allowed to increase, then an increase in $\rho_c$ leads to an increase in $\Delta T$, a trend noted previously \cite{KB24}. The reason for this is that at constant value $(T_\textrm{source}, T_\textrm{sink})$ and $\varepsilon_{LS}$, larger values $\sigma_{LS}$ lead to wider depletion layer thicknesses  $\delta_{LS}$. And since this depletion zone acts as a thermal insulation layer, the wider this zone, the larger the temperature drop $\Delta T$.
\begin{figure}[!htb]
\centering
\includegraphics{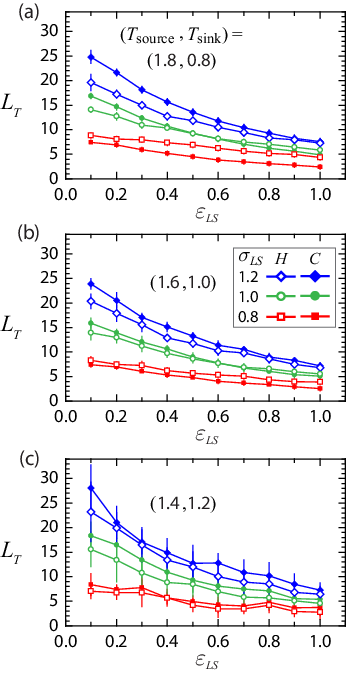}
\vspace{-0.1 in}
\small
\caption{(a) - (c) Reduction in the thermal slip length  $L_T$ at the hotter (H) and colder (C) L/S interface with increasing value $\varepsilon_{LS}$ and decreasing value $\sigma_{LS}$.}
\label{fig:LT}
\end{figure}
Shown in Fig. \ref{fig:LT} is the reduction in the thermal slip length $L_T $ evaluated at the hotter and colder interface with increasing $\varepsilon_{LS}$ and decreasing $\sigma_{LS}$. The larger fluctuations in Fig. \ref{fig:LT} (c) stem from the larger noise to signal ratio  of $\Delta T$ for $(T_\textrm{source}, T_\textrm{sink})=(1.4, 1,2)$ \cite{KT25}, as expected in cases subject to smaller temperature differentials  $T_\textrm{source} - T_\textrm{sink}$. For the same reasons mentioned above in regard to Fig. \ref{fig:Tjump}, here too the thermal slip length is not always larger at the colder L/S interface.

While historically physicists always measure the degree of thermal impedance at an L/S interface using the thermal slip length, researchers throughout different engineering communities still prefer to use the thermal boundary resistance defined as $\mathcal{R} = \Delta T/J_z$. For this study, we confirmed that the liquid (and solid) layers behave as Fourier media for which the magnitude of the thermal flux is then $J_z = k |dT/dz|_{liq}$ where $k$ is the effective thermal conductivity and $|dT/dz|_{liq}$ is a constant due to the linearity of the temperature profile $T_{liq}(z)$. As a result, the thermal boundary resistance is related to the thermal slip length through the simple relation  $\mathcal{R} = L_T/k$. Therefore, the data in Fig. \ref{fig:LT}, when multiplied by the factor $k^{-1}$, yield the corresponding curves for the thermal boundary resistance, which replicate the same curves as in Fig. \ref{fig:LT}, modulo the amplitude, and are therefore not reproduced here. The thermal conductivity values can be found in Ref. [\onlinecite{KT25}].
\begin{figure}[!htb]
\centering
\includegraphics{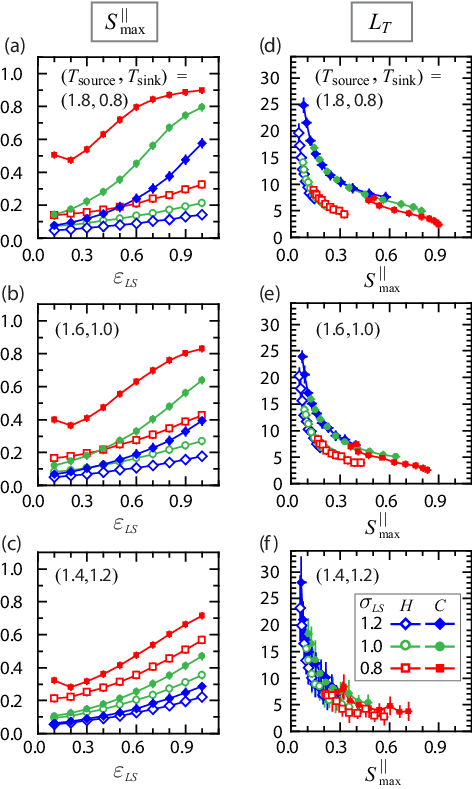}
\vspace{-0.1 in}
\small
\caption{(a)-(c) Increase in $S^\parallel_\textrm{max}$ measured at the hotter (H) and colder (C) L/S interface for decreasing temperature $(T_\textrm{source}$ or $T_\textrm{sink})$, increasing values $\varepsilon_{LS}$ or decreasing values $\sigma_{LS}$. (The three exceptions are discussed further in the text.) Connector lines are a guide to the eye. Legend in (f) applies to all six panels. (d)-(f) Reduction in thermal slip length $L_T$ with  increasing value $S^\parallel_\textrm{max}$.}
\label{fig:Smax}
\end{figure}

\subsection{Influence of long range translational order in contact layer}
Plotted in Fig. \ref{fig:Smax} (a) - (c) is the peak value of the in-plane structure factor $S^\parallel_\textrm{max}$, which all else equal, is always larger for particles in the colder contact layer, as expected. Its magnitude increases with decreasing $\sigma_{LS}$, increasing $\varepsilon_{LS}$ and decreasing local temperature enforced by lower temperatures  $T_\textrm{sink})$. Of the 180 systems represented, there are six special cases for which the structure factor at $\varepsilon_{LS} = 1.0$ exceeds the value  $S^\parallel_\textrm{max} > 0.8$, namely four cases with $T_\textrm{sink}=0.8$ and $\sigma=0.8$, and two cases with $T_\textrm{sink}=1.0$ and $\sigma=0.8$. (Additional information about these cases can be found in Ref. [\onlinecite{KT25}]). These cases also happen to exhibit near saturation in the values $S^\parallel_\textrm{max}$, due to strong binding with the solid lattice and formation of a solid-like contact layer. By contrast, all the other data exhibit a steady increase in $S^\parallel_\textrm{max}$ as $\varepsilon_{LS}$ increases from 0.1 to 1.0. The topmost curve in Fig. \ref{fig:Smax} (a)-(c) also reveals an interesting structural transition for $\varepsilon_{LS}=0.2$ and $\sigma_{LS}=0.8$, marked by the dip in $S^\parallel_\textrm{max}$ at $\varepsilon_{LS}=0.2$. This system represents the most disordered state of runs conducted with the coldest temperature $T_sink = 0.8$ and smallest value $\sigma_{LS}=0.8$. This combination of input variables leads to a more frustrated configuration of particles in the contact layer less able to adjust to the order and periodicity of the nearby solid lattice.

\subsection{Dependence of thermal slip length on long range translational order in contact layer}
Shown in Fig. \ref{fig:Smax} (d)-(f) is the reduction in the thermal slip length $L_T$ with increasing magnitude $S^\parallel_\textrm{max}$. This behavior confirms that a smaller thermal impedance is achieved when particles in the contact layer conform more closely to the order and periodicity of particles in the solid lattice. While this behavior is not unexpected, it is nonetheless interesting to note how rapid is the reduction in $L_T$ with increasing $S^\parallel_\textrm{max}$ and also how the data compress as the differential temperature $T_\textrm{source} - T_\textrm{sink}$ decreases. This behavior suggests the possibility of re-scaling the data onto a master curve, as described next.
\begin{figure}[!htb]
\centering
\includegraphics{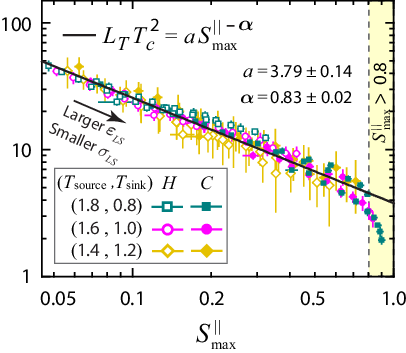}
\vspace{-0.1 in}
\small
\caption{Collapse of the data for the thermal slip length $L_T$ versus the in-plane structure factor of the contact layer when scaled by the contact layer temperature $T_c$. Solid curve is the best fit to Eq. (\ref{eqn:masterSmax}), excluding six points with $S_{max} > 0.8$ signifying  solid-like behavior.}
\label{fig:masterSmax}
\end{figure}

A nonlinear best fit to the power law relation
\begin{equation}
L_T \, T^2_c= a \, S^{-\alpha}_\textrm{max}~.
\label{eqn:masterSmax}
\end{equation}
was carried out using orthogonal distance regression so as to incorporate standard deviations in the measured values of $L_T$, $T_c$ and $S^\parallel_\textrm{max}$. The best fit, indicated by the superposed solid line in Fig. \ref{fig:masterSmax}, yielded an exponent $\alpha = 0.83 \pm 0.02$ and coefficient $a = 3.79 \pm 0.14$, where the values following $\pm$ denote 95\% confidence levels. To confirm this was the best fit, we conducted regression tests by reducing the exponent of $T_c$ from 2 to 1.5, which yielded $\alpha = 0.74 \pm 0.02$ and $a = 3.94 \pm 0.16$ and an increase in the residual sum of squares by about 20\%. We also tried fitting the data with $\alpha = 1.0$ and allowing the exponent of $T_c$ and the coefficient $a$ to vary; the exponent then increased to $2.34 \pm 0.12$ and the coefficient reduced to $3.10 \pm  0.11$ with an increase in the residual sum of squares by over 110\%. Allowing variation in all three fit constants $a$, $\alpha$ and the exponent of $T_c$ yielded a slight decrease in $\alpha$ from $0.83 \pm 0.02$ to $0.80 \pm 0.03$ and a decrease in the exponent of $T_c$ from 2 to $1.83 \pm 0.10$. Assuming rational exponents then, the analysis suggests a scaling relation of the form
\begin{equation}
L_T \sim  \frac{S^{\parallel 4/5}_\textrm{~max}}{T^2_c}~,
\label{eqn:powerlawA}
\end{equation}

\subsection{Dependence of thermal slip length on dominant vibrational frequencies in contact zone}
\label{sec:mastercurve}
\begin{figure}[!htb]
\centering
\includegraphics{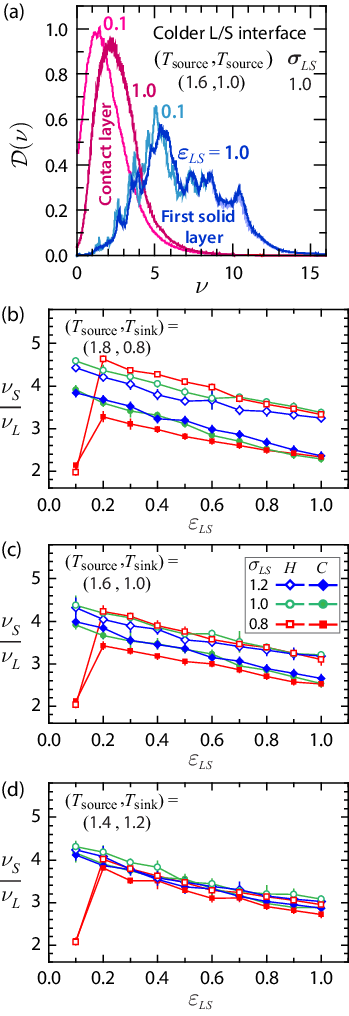}
\vspace{-0.1 in}
\small
\caption{(a) Vibrational frequency spectra per particle $\mathcal{D}(\nu)$ for the contact (L) layer and first solid (S) layer at the colder L/S interface for values $(T_\textrm{source}, T_\textrm{sink}) = (1.6,1.0)$, $\sigma_{LS}=1.0$ and $\varepsilon_{LS}=0.1~\textrm{and}~1.0$. Frequencies  $\nu_S$ and $\nu_L$ extracted from maxima of  $\mathcal{D}(\nu)$. (b) - (d) Reduction in ratio $\nu_S/\nu_L$ for larger values $\varepsilon_{LS}$ and colder contact layers. (Anomalous behavior for six data points with $\varepsilon_{LS}=0.1$ and $\sigma_{LS} = 0.8$ discussed in the text.)}
\label{fig:dos-freqratio}
\end{figure}

Given the master curve in $L_T$ showing strong correlation with the order and periodicity of the contact layer as induced by the crystalline surface potential, we also examined the vibrational spectra of particles in the contact (L) and first solid (S) layer. In particular, we computed the vibrational density of states $\mathcal{D}(\nu)$ given by Eq. (\ref{eqn:DOS}) for both layers and extracted the frequency ratio $\nu_S/\nu_L$, where $\nu_S$ and $\nu_L$ denote the frequencies corresponding to the respective maximum in the density of states. Shown in Fig. \ref{fig:dos-freqratio} (a) is an example of the density of states for the solid and liquid layers for the colder L/S interface obtained with  $(T_\textrm{source}, T_\textrm{sink}) = (1.6, 1.0)$, $\sigma = 1.0$ and $\varepsilon = 0.1$ and $1.0$. The data show that the density of states for the contact layer is more sensitive to the increase in $\varepsilon_{LS}$ than is the solid layer, as evident from the relatively larger shift in $\nu_L$ toward higher frequencies than occurs for $\nu_S$.

As well known, the value $\mathcal{D}(\nu = 0)$ is proportional to the self-diffusion coefficient of liquid particles. For an isotropic classical fluid of identical particles of mass $m$ at thermal equilibrium at temperature $T$, the self-diffusion coefficient $D = (k_B T/12\,m)~ \mathcal{D}(\nu=0)$ \cite{BW83}. While this relation is no longer exact for particles near a solid surface because of anisotropic effects like liquid layering, it is still the case that a smaller value $\mathcal{D}(\nu = 0)$ indicates a smaller diffusion coefficient. We confirmed that in systems at fixed values $(T_\textrm{source}, T_\textrm{sink})$ and $\sigma_{LS}$, $\mathcal{D}(\nu=0)$ decreased noticeably as $\varepsilon_{LS}$ was increased from 0.1 to 1.0, as evident in the example shown in Fig. \ref{fig:dos-freqratio} (a). As expected, an increase in the bonding strength $\varepsilon_{LS}$ hinders in-plane diffusion of liquid particles. A decrease in the in-plane mobility was also observed with colder contact layer temperatures $T_c$ enforced by lower reservoir temperatures  $T_\textrm{sink}$.

The data in Fig. \ref{fig:dos-freqratio} (b)-(d) confirm that smaller ratios $\nu_S/\nu_L$, indicative of stronger L/S vibrational coupling, is achieved with larger $\varepsilon_{LS}$, smaller $\sigma_{LS}$ and lower contact layer temperature by lowering $T_\textrm{source}$ or $T_\textrm{sink}$. The anomalous and large reduction in $\nu_S/\nu_L$ observed for all pairs $(T_\textrm{source}, T_\textrm{sink})$ with $\varepsilon_{LS}=0.1$ and $\sigma_{LS} = 0.8$ corresponds to those contact layers with $\nu_L \approx \nu_S/2$. Such behavior may indicate  formation of a constant layer with superlattice symmetry i.e. same symmetry as the [001] FCC solid facet but twice its lattice constant.
\begin{figure}[!htb]
\centering
\includegraphics{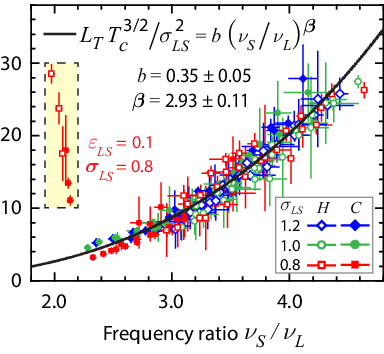}
\vspace{-0.1 in}
\small
\caption{Collapse of the data for the thermal slip length $L_T$ versus the frequency ratio $\nu_S/\nu_L$ when scaled by the contact layer temperature $T_c$ and LJ interaction distance $\sigma_{LS}$. Solid curve is the best fit to Eq. (\ref{eqn:masterfreq}). Fit excludes six data points for solid-like contact layers with superlattice symmetry, as  described in the text.}
\label{fig:masterfreq}
\end{figure}
Shown in Fig. \ref{fig:masterfreq} is the collapse of the data for the thermal slip length in terms of the dominant vibration frequencies in the L/S contact zone when scaled by the contact layer temperature and LJ interaction distance $\sigma_{LS}$. Clearly, the six cases just noted with contact layers resembling solid-like super lattices  don't fit the general trend. The best fit to the data, indicated by the solid line, represents a nonlinear fit to the relation
\begin{equation}
L_T \, T^{3/2}_c/\, \sigma^2_{LS}=
b \, \bigg(\frac{\nu_S}{\nu_L}\bigg)^\beta~.
\label{eqn:masterfreq}
\end{equation}
obtained by orthogonal distance regression, which incorporates the standard deviations in the measured values of $L_T$, $T_c$ and $\nu_S/\nu_L$. The best fit yielded  $\beta = 2.93 \pm 0.11$ and $b=0.35 \pm 0.05$, where the values following $\pm$ denote 95\% confidence levels. Expanding the search to allow variation of all the variables shown resulted in little change to the exponent of $\sigma_{LS}$, a small increase in the exponent of $T_c$ from $3/2$ to $1.61 \pm 0.13$ and an even smaller increase in the exponent $\beta$ from $2.93 \pm 0.11$ to $2.98 \pm 0.14$. Regression attempts based on third order polynomials in $\nu_S/\nu_L$ led to substantially worse fits no matter the initial seed values.

Assuming power law behavior with rational exponents, the analysis for the thermal slip length behavior in the frequency domain suggests the scaling relation
\begin{equation}
L_T \sim \frac{\sigma^2_{LS}}{T^{3/2}_c} \bigg(\frac{\nu_S}{\nu_L} \bigg)^3~.
\label{eqn:powerlawB}
\end{equation}

\section{Conclusion}
This computational study was designed to elicit the dependence of the thermal slip length on \textit{correlated  behavior} within the L/S contact zone comprising the first solid and first liquid layer, the latter called the contact layer. In particular, we focused on the long range translational order and vibrational spectrum in the contact layer as influenced by the order and symmetry of the solid crystal. For the layered system used to examine the thermal transfer process, the input parameters were restricted to the temperatures of the Langevin source and sink reservoirs and the Lennard-Jones intermolecular parameters $\varepsilon_{LS}$ and $\sigma_{LS}$. Different parameter sets naturally generated different values of the thermal flux propagating across the S/L/S system and different contact layer temperatures. Data was collected simultaneously from the hotter and colder L/S interface yielding a total of 180 systems. The two quantities used to measure correlated behavior across the L/S interface were the peak value of the in-plane structure factor of the contact layer, which coincided with the  smallest set of reciprocal lattice vectors of the solid surface, and the ratio of frequencies defined by the maxima in the vibrational density of states of the first solid and contact layer.

Excluding a handful of cases involving solid-like and not liquid-like contact layers, the data for the thermal slip length $L_T$ versus structure factor, when scaled by a power of $T_c$, collapse nicely onto a master curve described by a simple power law relation. Likewise in the frequency domain, the data for $L_T$ also undergo collapse onto a simple power law equation when scaled by (different) powers of $T_c$ and $\sigma_{LS}$. From the well known property of corresponding states applicable to the Lennard-Jones potential, we anticipate similar power law relations for other LJ systems so long as the fluid state is not near a critical or triple point. In future studies, it will be interesting to explore how more complex intermolecular potentials for L/S interactions modify these  power law relations.

The important takeaways from this work are not the  numerical values of the fit constants but the fact that the thermal slip length is governed by power law behavior stemming from the correlated spatial and frequency behavior in the L/S contact zone. More generally, these power law relations underscore the critical role of \textit{surface localized phonons} in regulating thermal flux. While it has been known for decades that surface phonons from diffusive scattering at a rough interface are responsible for the fact that the thermal boundary resistance of a helium liquid/helium solid interface is much smaller \cite{N85} than the value predicted by the acoustic mismatch theory \cite{K52}, we know of no such studies related to surface phonon localization in normal L/S interfaces.

On the experimental side, L/S interfaces pose many more challenges to measurement than S/S interfaces, and so validation of the relations reported in this work is likely not imminent. Perhaps high-resolution techniques combining small angle neutron scattering with neutron reflectometry may be successful in extracting measurements of the  in-plane static structure factor and vibrational spectrum of the contact layer for systems in thermal non-equilibrium. The more difficult challenge, which has persisted for decades, will be accurate measurement of the temperature drop in normal (i.e. non superfluid) L/S systems, which may require new instrumentation. For these reasons, we still view particle based simulations as the only viable high resolution probe for quantifying the thermal boundary impedance of a normal L/S interface.
\section{Data Availability Statement}
The data for this study are publicly available \cite{KT25}.
\vspace{0.1in}
\noindent
\vspace{-0.1in}
\begin{acknowledgments}
\vspace{-0.1 in}
The authors gratefully acknowledge funding from a 2019 NASA Space Technology Research Fellowship (HK) and the generous assistance of Dr. Peter Thompson, IT administrator of the  \texttt{{LIS2T}} computing cluster used in this study.
\end{acknowledgments}
\vspace{-0.3in}
\section*{Author Contribution Statement}
Conceptualization: SMT; Data curation: SMT;
Formal analysis: HK, SMT; Funding acquisition: SMT; Investigation: HK, SMT; Methodology:
SMT; Project administration: SMT; Resources:
SMT; Software: HK, SMT; Supervision: SMT;
Validation: HK, SMT; Visualization: HK, SMT;
Writing - original draft: SMT; Writing - review
\& editing: SMT


\begin{thebibliography}{58}
\bibitem [{FN1()}]{FN1new}%
  \BibitemOpen
  \href@noop {} {}\bibinfo {note} {In many studies of the L/S interface, the
  liquid structure factor is often computed from the relation
  $|\sum_{p}^{N_c}\exp (-i \bm{k}\cdot\bm{r}_p)|^2$, a simplification strictly
  valid for particles on sites of a Bravais lattice.}\BibitemShut {Stop}%
\end{thebibliography}
\end{document}